\documentclass[preprint]{elsarticle}

\usepackage{lineno}
\usepackage{hyperref}
\hypersetup{
    colorlinks=true,
    linkcolor=black,
    citecolor=black,
    filecolor=black,
    urlcolor=black,
}
\usepackage{amsmath}
\usepackage{amssymb}
\usepackage{verbatim}
\usepackage{epstopdf}
\makeatletter
\def\ps@pprintTitle{%
   \let\@oddhead\@empty
   \let\@evenhead\@empty
   \def\@oddfoot{\reset@font\hfil\thepage\hfil}
   \let\@evenfoot\@oddfoot
}
\def\onedot{$\mathsurround0pt\ldotp$}
\def\cddot{
  \mathbin{\vcenter{\baselineskip.67ex
    \hbox{\onedot}\hbox{\onedot}}%
  }}%
\makeatother
\modulolinenumbers[5]

\journal{}

\newcommand{\hhat}[1]{\hat{\hat{#1}}}

\begin{document}

\begin{frontmatter}

\title{A FFT-based finite-difference solver for massively-parallel direct numerical simulations of turbulent flows\tnoteref{mytitlenote}}
\tnotetext[mytitlenote]{The numerical tool is freely available in \href{http://github.com/p-costa/CaNS}{\color{black}github.com/p-costa/CaNS}.}

\author{Pedro Costa\fnref{myfootnote}}
\address{KTH, Department of Mechanics, SE-100 44 Stockholm, Sweden}
\ead{pedrosc@mech.kth.se}




\begin{abstract}
We present an efficient solver for massively-parallel direct numerical simulations of incompressible turbulent flows. The method uses a second-order, finite-volume pressure-correction scheme, where the pressure Poisson equation is solved with the method of eigenfunction expansions. This approach allows for very efficient FFT-based solvers in problems with different combinations of homogeneous pressure boundary conditions. Our algorithm explores all combinations of pressure boundary conditions valid for such a solver, in a single, general framework. The method is implemented in a 2D \emph{pencil}-like domain decomposition, which enables efficient massively-parallel simulations. The implementation was validated against different canonical flows, and its computational performance was examined. Excellent strong scaling performance up to $10^4$ cores is demonstrated for a domain with $10^9$ spatial degrees of freedom, corresponding to a very small wall-clock time/time step. The resulting tool, \emph{CaNS}, has been made freely available and open-source.
\end{abstract}

\begin{keyword}
Direct Numerical Simulations, Turbulent Flows, High-Performance Computing, Fast Poisson Solver
\end{keyword}

\end{frontmatter}


\section{Introduction}
Turbulent flows are ubiquitous in nature and industry, being the most common flow regime for cases which dimensions the human eye can depict. These flows exhibit unsteady, three-dimensional, chaotic and multi-scale dynamics.  The Navier-Stokes equations governing its dynamics are highly non-linear, making analytical predictions often difficult. Fortunately, the continuous increase in computer power, together with the progress in development of efficient numerical techniques, resulted in a paradigm change in turbulence research. It is now possible to conduct direct numerical simulations (DNS) which generate data for the full spectrum of scales with billions, or even trillions of spatial degrees of freedom \cite{Ishihara-et-al-ARFM-2009}.\par
Pseudo-spectral approaches have been the method of choice for DNS of relatively simple flows, like homogeneous isotropic turbulence \cite{Orszag-and-Patterson-PRL-1972} or pressure-driven wall-bounded flows \cite{Kim-et-al-JFM-1987,Gavrilakis-JFM-1992}. In these cases one can benefit from the spectral spatial convergence of the solution, and explore the FFT algorithm for efficient computations. In many situations, however, lower-order discretizations are desirable. Finite-difference methods, for instance, can reproduce several lower-order moments of a flow observable with the same fidelity of a spectral method, and much less computational effort \cite{Vreman-and-Kuerten-PoF-2014,Kooij-et-al-CF-2018} (despite requiring more resolution, and consequently higher memory bandwidth per FLOP). Moreover, these methods are in general more versatile in terms of boundary conditions and problem complexity \cite{Orlandi-2012}. For instance, in finite-difference algorithms, complex geometries can be easily and efficiently implemented through immersed-boundary methods \cite{Fadlun-et-al-JCP-2000,Mittal-and-Iccarino-ARFM-2005}, without facing problems due to Gibbs phenomenon.\par
Indeed, many fundamental insights on the dynamics of turbulent flows have been revealed with standard second-order finite-difference algorithms embedded in a pressure-correction scheme. Few of many examples are the turbulent channel flow with rough walls \cite{Leonardi-et-al-FFM-2003}, flow through porous media \cite{Breugem-et-al-JFM-2006}, interface-resolved simulations of particle-laden flows \cite{Uhlmann-et-al-JCP-2005}, and turbulent Rayleigh-B\'enard convection \cite{Verzicco-and-Camussi-JFM-2003}. This class of methods allows for fast computations, and is often used e.g.\ when a dense disposition of immersed boundaries reduces anyway the overall accuracy of the method to second-order. \par
The Poisson equation for the correction pressure is often the most demanding part of the Navier-Stokes solver. Consider the second-order finite-difference discretization of the Laplacian operator. In many cases, iterative solvers (e.g.\ multigrid methods \cite{Golub-et-al-2012}) are used to solve the resulting system. These methods exhibit excellent scaling properties and are versatile when it comes to the type of boundary conditions that can be implemented. However, much more efficient direct solvers can be used for several combinations of homogeneous pressure boundary conditions. The method of eigenfunction expansions \cite{Wilhelmson-and-Ericksen-JCP-1977,Swarztrauber-1977,Schumann-and-Sweet-JCP-1988} is an example. This approach has been implemented in the 1980s in the well-known \texttt{FISHPAK} library \cite{FISHPAK} and explores the fact that the number of diagonals in the coefficient matrix can be reduced by solving an eigenvalue problem with Fourier-based expansions. When compared to multigrid methods, FFT-based direct solvers for the second-order finite-difference Poisson equation can be $O(10)$ times faster \cite{Dodd-and-Ferrante-JCP-2014}, and by construction satisfy the solution to machine precision.\par
Consider, for instance, a 3D problem with at least two periodic boundary conditions. Using the method of eigenfunction expansions, one can apply a discrete Fourier transform (DFT) operator in two directions, and solve a resulting tridiagonal system with Gauss elimination. The inherent challenge for an efficient parallelization is that the FFT algorithm requires all the points in one direction. Consequently, a distributed-memory parallelization in more than one direction is not straightforward. This difficulty has restricted the progress in scaling these methods to a very high number of cores (e.g.\ more than $O(10^3)$).\par 
There has been a change in trend since highly-scalable libraries for two-dimensional \emph{pencil-like} domain decomposition \cite{Li-and-Laizet-CRAY-2010} started to appear. Several recent examples of numerical implementations using this direct method, combined with a 2D domain decomposition, achieved unprecedented performances in problem sizes with $O(10^9)$ grid points. Examples are turbulent Taylor-Couette flows \cite{Ostilla-et-al-JFM-2016}, interface-resolved simulations of bubbly flows in homogeneous isotropic turbulence \cite{Dodd-and-Ferrante-JFM-2016}, and interface-resolved simulations of turbulent wall-bounded suspensions \cite{Costa-et-al-PRL-2016}.\par
Recently, \citet{vanderPoel-et-al-CF-2015} published a numerical algorithm\footnote{Freely available in \href{http://github.com/PhysicsofFluids/AFiD}{\color{black}github.com/PhysicsofFluids/AFiD}.} for wall-bounded turbulence that showed very good performance both in terms of scaling, up to $64\,000$ cores, and in terms of actual wall-clock time. In their work and the studies mentioned in the previous paragraph, discrete Fourier transforms (DFT) could be applied to solve the second-order finite-difference Poisson equation, as there were at least two periodic directions. However, other types of FFT-based expansions can be used for several combinations of homogeneous pressure boundary conditions \cite{Schumann-and-Sweet-JCP-1988,Orlandi-2012,Fuka-AMC-2015}. 
To the best of our knowledge, there is no general implementation for massively-parallel DNS that features the wide range of boundary conditions that can be covered with a second-order, FFT-based solver. This is the main motivation of the present work.\par
We present a numerical method for fast, massively-parallel numerical simulations of turbulent flows. The corresponding code, \emph{CaNS} (Canonical Navier-Stokes), benefits from the efficiency of FFT-based codes for the finite-difference discretization of the pressure Poisson equation, and allows for simulating a wide range of canonical flows.\par
The manuscript is organized as follows. In section~\ref{sec:numerics} we describe the method and provide some mathematical background on the Poisson solver. Section~\ref{sec:implementation} describes details on the implementation of the numerical algorithm for massively-parallel numerical simulations of turbulent flows. Section~\ref{sec:results} presents a validation of the code for distinct flows, and accesses its computational performance in thousands of cores. Finally, in section~\ref{sec:conclusions} we conclude and discuss future perspectives.

\section{Numerical Method} \label{sec:numerics}
The numerical algorithm solves the Navier-Stokes equations for an incompressible, Newtonian fluid with unit density, and kinematic viscosity $\nu$,
\begin{subequations}
\begin{align}
\boldsymbol{\nabla} \cdot \mathbf{u} = 0 \mathrm{,} \label{eqn:cont} \\
\frac{\partial \mathbf{u}}{\partial{t}} + (\mathbf{u} \cdot \boldsymbol{\nabla})\mathbf{u} = -\boldsymbol{\nabla}{p} + \nu\boldsymbol{\nabla}^2\mathbf{u}\mathrm{,} \label{eqn:mom}
\end{align} 
\end{subequations}
where $\mathbf{u}$ and $p$ are the fluid velocity vector and pressure, respectively.\par
These equations are solved in a structured Cartesian grid, uniformly-spaced in two directions. Standard second-order finite-differences are used for spatial discretization with a staggered (marker and cell) disposition of grid points. The equations above are coupled through a pressure-correction method \cite{Amsden-and-Harlow-JCP-1970}, and integrated in time with a low-storage, three-step Runge-Kutta scheme (RK3) \cite{Wesseling-2009}. The advancement at each substep $k$ is presented below in semi-discrete notation ($k=1,2,3$; $k=1$ corresponds to a time level $n$ and $k=3$ to $n+1$):
\begin{subequations}
\begin{align}
\mathbf{u}^* = \mathbf{u}^k + \Delta t\left(\alpha_k\mathbf{AD}^{k} + \beta_k\mathbf{AD}^{k-1} - \gamma_k\boldsymbol{\nabla} p^{k-1/2}\right)\mathrm{,} \label{eqn:up} \\
\nabla^2\Phi = \frac{\boldsymbol{\nabla}\cdot\mathbf{u}^*}{\gamma_k\Delta t}\mathrm{,}\label{eqn:poi_ns} \\
\mathbf{u}^k = \mathbf{u}^* - \gamma_k\Delta t \boldsymbol{\nabla}\Phi\mathrm{,} \\
p^{k+1/2} = p^{k-1/2} + \Phi\mathrm{,}
\end{align}
\end{subequations}
where $\mathbf{AD}\equiv-(\mathbf{u}\cdot\boldsymbol{\nabla})\mathbf{u}+\nu\nabla^2\mathbf{u}$, $\mathbf{u}^*$ is the prediction velocity and $\Phi$ the correction pressure. The RK3 coefficients are given by $\alpha_k=\lbrace{8/15,5/12,3/4\rbrace}$, $\beta_k=\lbrace{0,-17/60,-5/12\rbrace}$ and $\gamma_k=\alpha_k+\beta_k$. This temporal scheme has been proven to be reliable for DNS of turbulent flows, yielding overall second-order temporal and spatial accuracy \cite{Verzicco-and-Orlandi-JCP-1996}.\par
A sufficient criterion for a stable temporal integration is given in \cite{Wesseling-2009}:
\begin{equation}
\Delta t < \min\left(\frac{1.65\Delta r^2}{\nu},\frac{\sqrt{3}\Delta r}{\max_{ijk}(|u|+|v|+|w|)}\right)\mathrm{,}
\end{equation}
with $\Delta r = \min(\Delta x,\Delta y,\Delta z)$ and $\Delta {x_i}$ the grid spacing in direction $x_i\equiv\lbrace{x,y,z\rbrace}$. The time step restriction due to the viscous effects can be removed with an implicit discretization (e.g.\ Crank-Nicolson) of the diffusion term. This involves the solution of three Helmholtz equations, one for each component of the prediction velocity. The associated computational overhead can pay off in case of flows at low Reynolds number, but is unnecessary for the inertia-dominated flows of our interest.\par
Finally, let us note that the boundary conditions for the correction pressure and prediction/final velocity may not be specified independently. For instance, for a prescribed velocity, the pressure boundary condition must be set to homogeneous Neumann (i.e.\ zero gradient in the direction normal to the boundary), such that the projection step does not alter this condition.
\subsection{Poisson Equation}
Often the solution of the Poisson equation for the correction pressure, Eq.~\eqref{eqn:poi_ns},  is the most computation-intensive part of a Navier-Stokes solver. Even so, there are several configurations for which a fast (FFT-based), direct method can be used, even if the unknown is non-periodic. To achieve this, one can explore the method of eigenfunction expansions, as in \cite{Schumann-and-Sweet-JCP-1988}. This method is applied in two domain directions, $x$ and $y$, requiring therein a constant grid spacing, and homogeneous boundary conditions. Since this part of the algorithm comprises the most elaborate parallelization steps, we will briefly introduce some mathematical background below.\par
Consider the following constant-coefficients Poisson equation discretized with second-order central differences at grid cell $i,j,k$:
\begin{align}
  &(\Phi_{i-1,j  ,k  }-2\Phi_{i,j,k}+\Phi_{i+1,j  ,k  })/\Delta x^{2} + \nonumber\\
  &(\Phi_{i  ,j-1,k  }-2\Phi_{i,j,k}+\Phi_{i  ,j+1,k  })/\Delta y^{2} + \nonumber\\
  &(\Phi_{i  ,j  ,k-1}-2\Phi_{i,j,k}+\Phi_{i  ,j  ,k+1})/\Delta z^{2} = f_{i,j,k}\mathrm{;}\label{eqn:poi}
\end{align}
The method reduces this system of equations with $7$ non-zero diagonals to a tridiagonal system, which can be solved very efficiently with Gauss elimination. To achieve this we apply a discrete operator $\mathcal{F}^{x_i}$ to Eq.~\eqref{eqn:poi} in two domain directions, that reduces the problem to:
\begin{align}
  (\lambda_{i}/\Delta x^2 + \lambda_{j} /\Delta y^2)\hhat{\Phi}_{i,j,k} +
  (\hhat{\Phi}_{i,j,k-1}-2\hhat{\Phi}_{i,j,k}+\hhat{\Phi}_{i,j,k+1})/\Delta z^{2} = \hhat{f}_{i,j,k}\mathrm{,} \label{eqn:poi_reduced}
\end{align}
where $\hhat{\square}\equiv\mathcal{F}^y(\mathcal{F}^x(\square))$ and, although not necessary, we consider $\Delta z$ constant for simplicity\footnote{In the numerical tool, the grid spacing in $z$ can be non-uniform.}. The eigenvalues $\lambda$ and operators $\mathcal{F}^{x_i}$ depend on the boundary conditions of the problem, which need to be satisfied by the corresponding inverse (backward) operator, $\mathcal{F}^{-1\,x_i}$ (e.g., for a periodic boundary condition $\mathcal{F}$ is the Discrete Fourier Transform (DFT)). For several non-periodic combinations of boundary conditions, $\mathcal{F}$ corresponds to well-known discrete transforms (DT) that can be expressed in terms of DFT; see \cite{Schumann-and-Sweet-JCP-1988,Orlandi-2012}. This allows for efficient FFT algorithms with little computational overhead with respect to the periodic case. An operator could in principle be applied a third time, in direction $z$. However, the tridiagonal system in Eq.~\eqref{eqn:poi_reduced} is solved more efficiently with Gauss elimination: $O(n)$ operations, contrasting with $O(n\log n)$ required by the FFT algorithm, with $n$ being the number of grid cells in $z$. Moreover, a non-uniform grid spacing in $z$ is possible with Gauss elimination.\par
Tables~\ref{tbl:operators} and~\ref{tbl:transfs} summarize the transforms pertaining to different boundary conditions. Since the pressure grid cells in the Navier-Stokes solver are staggered, we solely present the transforms whose inverse satisfy homogeneous staggered boundary conditions. Note, however, that non-staggered versions must be used when the temporal integration of the diffusion term in Eq.~\eqref{eqn:up} is implicit.
\begin{table}[!htbp]
\centering
\caption{Eigenvalues, and forward ($\mathcal{F}$) and backward ($\mathcal{F}^{-1}$) transforms for different combinations of boundary conditions. P, D and N denote respectively periodic, and staggered Dirichlet and Neumann boundary conditions (BC). The eigenvalues (Eq.~\ref{eqn:poi_reduced}) are given by $\lambda_q = -4\sin^2(\theta_q)$, $q=0,1,\dots,n-1$ \cite{Schumann-and-Sweet-JCP-1988}; $p=0,1,\dots,n/2-1$ and $n$ is the (even) number of grid cells in the one direction. The mathematical expressions for $\mathcal{F}$ are shown in Table~\ref{tbl:transfs}.}\label{tbl:operators}
\def\arraystretch{2.5}
\begin{tabular}{ c c c c c }
\hline
\hline
  BC            & $\theta_q$               &  $\mathcal{F}      $  & $\mathcal{F}^{-1}$ \\
\hline
\hline
  P-P           &  $\begin{cases}\displaystyle\frac{(p+1)\pi}{n}, q=2p+1 \\
                                 \displaystyle\theta_{p-1} ~~~~~~~ , q=2p \ne 0   \\
                                 \displaystyle 0  ~~~~~~~~~~~~    , q = 0 \end{cases}$                    &  DFT            & $\displaystyle\frac{1}{n }$~IDFT    \\
\hline
  N-N           & $\displaystyle\frac{ q    \pi}{2n}$   &  DCT-II         & $\displaystyle\frac{1}{2n}$~DCT-III \\
\hline
  D-D           & $\displaystyle\frac{ (q+1)  \pi}{2n}$   &  DST-II         & $\displaystyle\frac{1}{2n}$~DST-III \\
\hline
  N-D           & $\displaystyle\frac{ (2q+1)   \pi}{4n}$   &  DCT-IV         & $\displaystyle\frac{1}{2n}$~DCT-IV \\
\hline
\hline
\end{tabular}
\end{table}
\begin{table}
\centering
\caption{Coefficients $\hat{a}_q$, ($q=0,1,\dots,n-1$) for a discrete transform $\hat{a} = \mathcal{F}(a)$ of a sequence of $n$ real numbers $a\equiv\lbrace{a_0,a_1,\dots,a_{n-1}\rbrace}$ \cite{Schumann-and-Sweet-JCP-1988}. $n$ is assumed to be even and $l=0,1,\dots,n/2-1$.}\label{tbl:transfs}
\def\arraystretch{1.5}
\begin{tabular}{ c l }
  \hline\hline
  \multicolumn{1}{c}{$\mathcal{F}      $} & \multicolumn{1}{c}{$\hat{a}_q$} \\
  \hline\hline
   DFT    & $\begin{cases}
              \sum_{p=0}^{n-1}a_p~~~~~~~~~~~~~~~~~~~~ ,~q      = 0                   \\
              \sum_{p=0}^{n-1}a_p\cos( 2\pi pl/n)     ,~q = 2l+1 \ne n-1\\
              \sum_{p=0}^{n-1}a_p\sin( 2\pi pl/n)     ,~q = 2l   \ne 0                 \\
              \sum_{p=0}^{n-1}a_p(-1)^p~~~~~~~~~~~    ,~q        = n-1                 \\
  \end{cases}$
\\
  \hline
   IDFT   & $a_0  +
             \sum_{p=0}^{n/2-2}(a_{2p+1}\cos(2\pi pq/n) + a_{2p+2}\sin(2\pi pq/n))~+
             a_{n-1}(-1)^q    $  \\
  \hline
   DCT-II  & $2\sum_{p=0}^{n-1}a_p\cos(\pi(p+1/2)q/n)$              \\ 
  \hline
   DCT-III & $a_0+2\sum_{p=1}^{n-1}a_p\cos(\pi p(q+1/2)/n)$         \\ 
  \hline
   DST-II  & $2\sum_{p=0}^{n-1}a_p\sin(\pi(p+1/2)(q+1)/n)$                \\ 
  \hline
   DST-III & $2\sum_{p=0}^{n-2}a_p\sin(\pi (p+1)(q+1/2)/n) + (-1)^{q}a_{n-1}$ \\ 
  \hline
   DCT-IV  & $2\sum_{p=0}^{n-1}a_p\cos(\pi (p+1/2)(q+1/2)/n)$         \\ 
  \hline\hline
\end{tabular}
\end{table}
\par Finally, the steps required for solving the Poisson, and associated number of operations are shown below (with $N_{x_i}$ the number of points in direction $x_i$):
{
\begin{itemize}
    \item[1:] compute $N_z$ times the discrete forward transform of $f$ in directions $x$ and $y$ successively, $\hhat{f}=\mathcal{F}^y(\mathcal{F}^x(f))\,\rightarrow$ $\mathcal{O}(N_z(N_yN_x\log N_x+N_xN_y\log N_y))$ operations;
    \item[2:] solve the resulting $N_xN_y$ tridiagonal systems for the pressure with Gauss elimination -- $\mathcal{O}(N_xN_yN_z)$ operations;
    \item[3:] compute $N_z$ times the discrete backward transform of $\hhat{f}$ in directions $y$ and $x$ successively, $f=\mathcal{F}^{-1,x}(\mathcal{F}^{-1,y}(\hhat{f}))\,\rightarrow$ $\mathcal{O}(N_z(N_xN_y\log N_y+N_yN_x\log N_x))$ operations.
\end{itemize}
}

\section{Implementation}\label{sec:implementation}
\begin{figure}[!htbp]
  \includegraphics[width=0.32\textwidth]{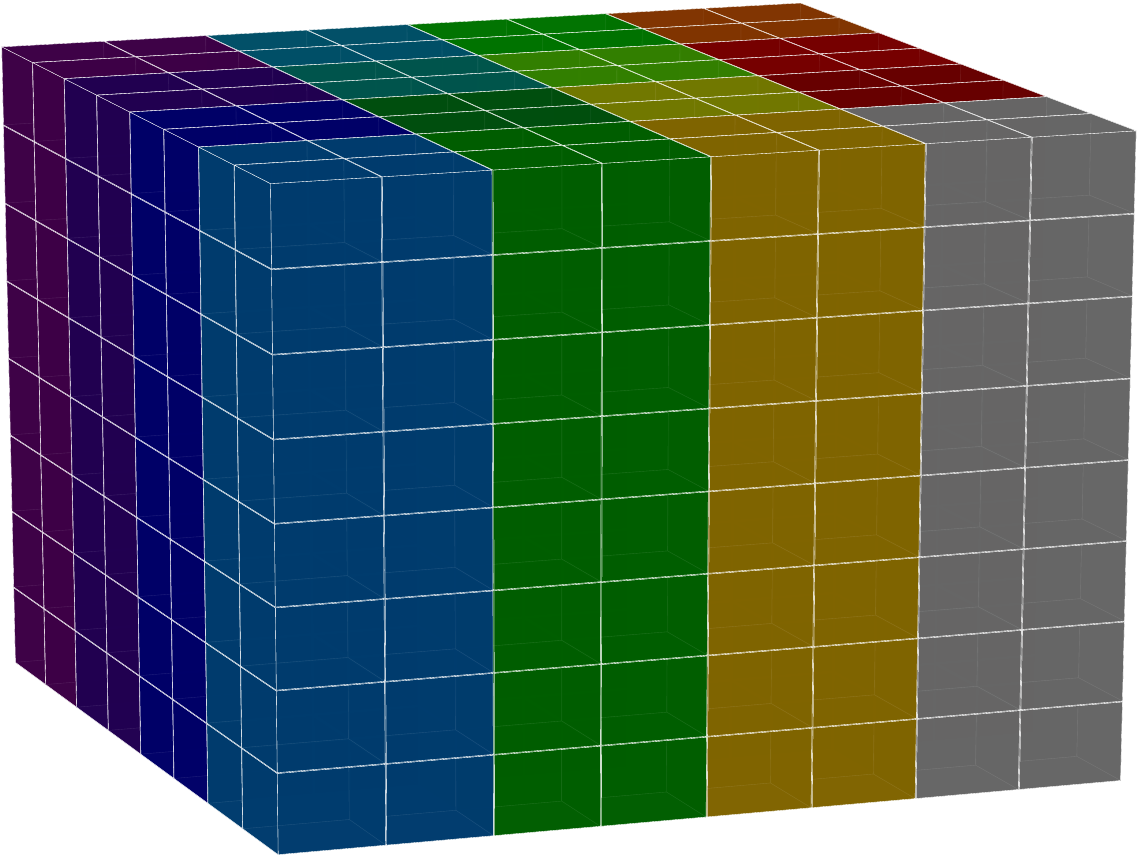}\hfill
  \includegraphics[width=0.32\textwidth]{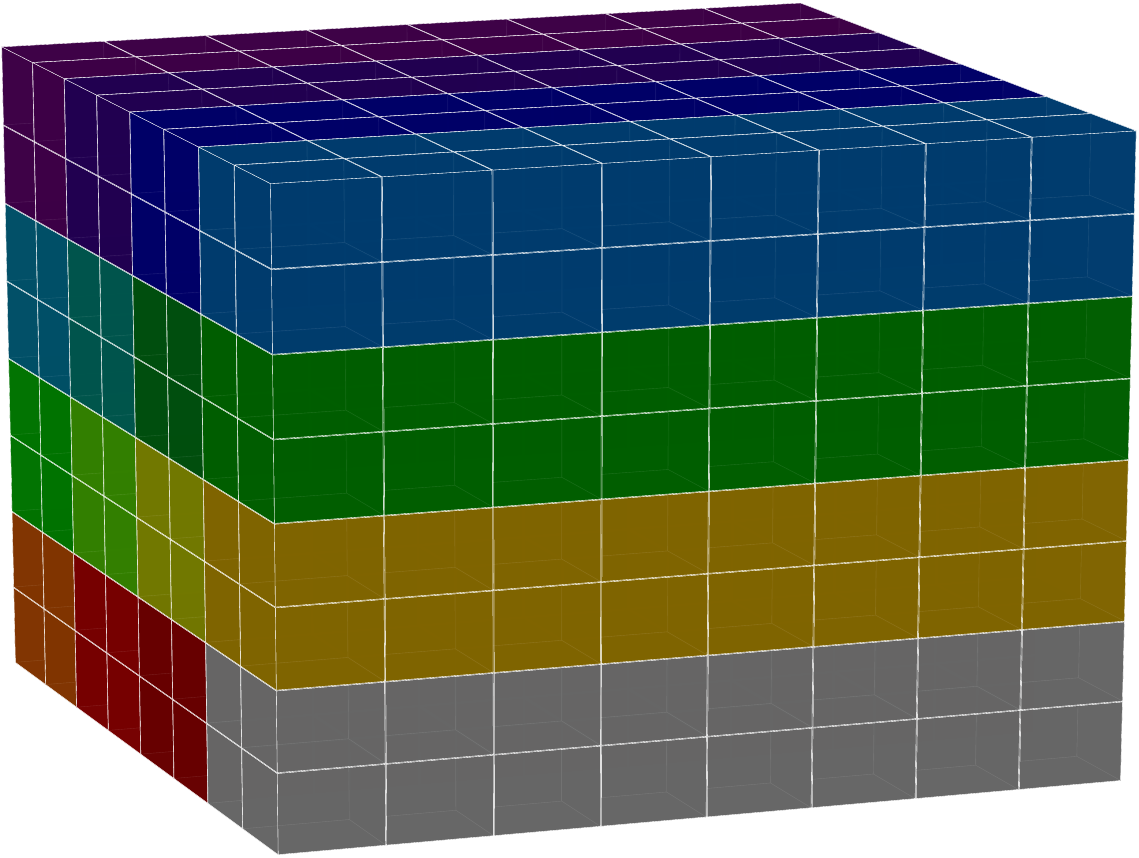}\hfill
  \includegraphics[width=0.32\textwidth]{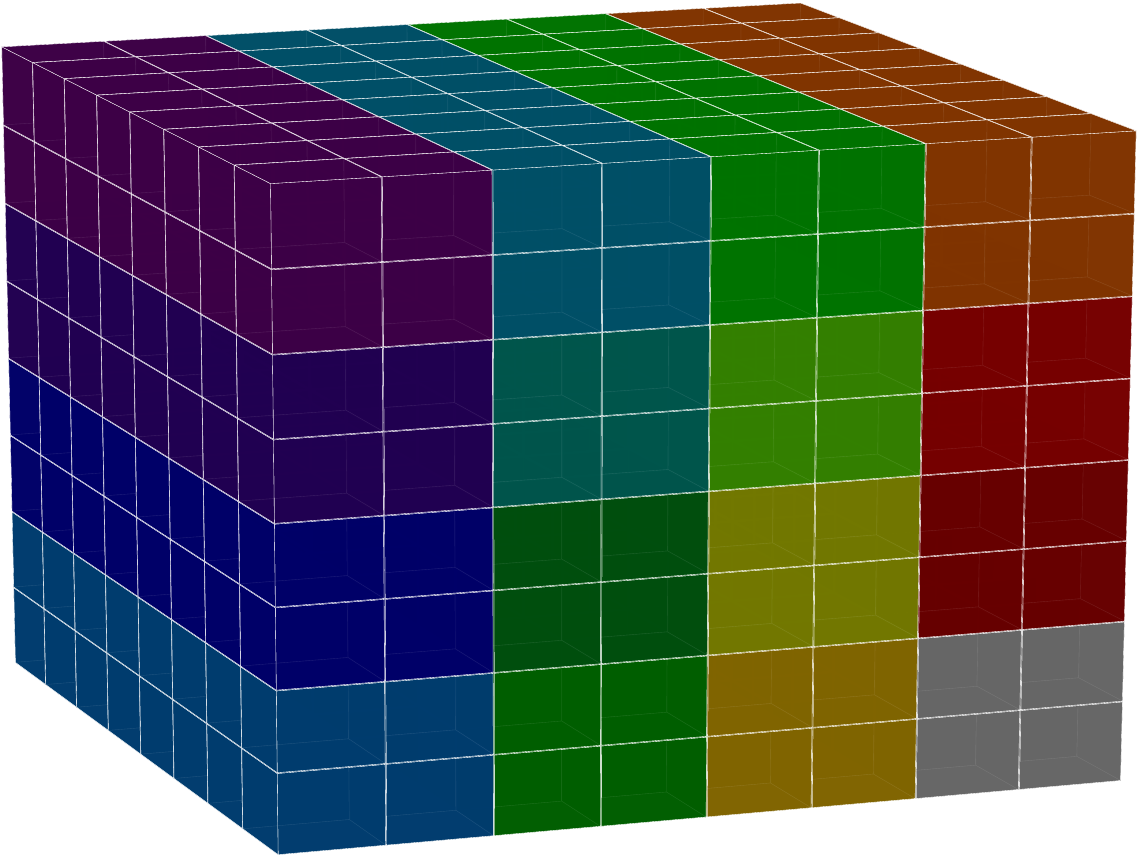}\hfill
  \put(-280,-5){$x$}
  \put(-336, 4){$y$}
  \put(-352,45){$z$}
  \caption{Illustration of a pencil-like domain decomposition, with $N_x \times N_y \times N_z = 8^3$ grid cells and $N_{p}^x \times N_{p}^y = 4^2$ computational subdomains (color-coded). The left-most panel corresponds to the configuration used throughout most steps of the algorithm. Transpositions of the data distribution to $x$-aligned (middle) and $y$-aligned (right) pencils are performed in the Poisson solver to compute efficiently the discrete transforms $\mathcal{F}^{x_i}$.}\label{fig:2decomp}
\end{figure}

The numerical algorithm is implemented in \texttt{FORTRAN90/95}, with a Message-Passing Interface (\texttt{MPI}) extension for distributed-memory parallelization, combined with a shared-memory parallelization (hybrid \texttt{MPI-OpenMP}). The geometry is divided into several computational subdomains in a \textit{pencil}-like decomposition (figure~\ref{fig:2decomp}). Throughout most steps of the algorithm, $N_{p}^x \times N_{p}^y$ pencils are aligned in the $z$ direction.\par
The $2$-cell width of the finite-difference discretization requires communication. Following common practice, we use \emph{halo} cells that store a copy of data pertaining to the boundary of an adjacent subdomain. This requires four pairwise data exchanges (e.g.\ \texttt{SEND\_RECV}) per halo update, and has negligible computational overhead.\par
Further communication steps are required for computing the DT: the operators are applied successively in directions $x$ and $y$, requiring all the grid cells for the direction considered. A simple solution would be to change the distribution from 2D to a 1D \textit{slab}-like configuration, decomposing the domain only in $z$. This requires a single \texttt{MPI\_ALL\_TO\_ALL} operation, but restricts the number of slabs to $N_{p}^x N_{p}^y \leq N_z$. Combining this approach with a shared-memory parallelization (e.g.\ hybrid \texttt{MPI-OpenMP}) relaxes the restriction by a factor $O(10)$ for the present state-of-the-art present hardware (see e.g.\ \href{https://www.top500.org/}{\color{black}top500.org}), which can still be constraining for relatively large problem sizes, e.g.\ with $N_z=O(10^2)$ and very large $N_xN_y$.\par
This issue can be circumvented by keeping the 2D decomposition, and transposing the data distribution such that it is shared in the direction of interest, as in figure~\ref{fig:2decomp}. This can be done at the cost of a one extra \emph{all-to-all} operation per DT. To achieve this we use the highly-scalable \texttt{2DECOMP\&FFT} library \cite{Li-and-Laizet-CRAY-2010}. This library provides a simple interface to transpose the decomposition from $x$ to $y$-aligned pencils, from $y$ to $z$, and the reciprocal operations $y$ to $x$ and $z$ to $y$. The authors of the \texttt{AFiD} code \cite{vanderPoel-et-al-CF-2015} implemented a direct transposition from $x$-aligned to $z$-aligned pencils and vice-versa, which was not present in the original \texttt{2DECOMP\&FFT} library and avoids one extra \emph{all-to-all} operation when transposing from $x$ to $z$. We ported these modifications to our DNS code, although in the present study we still use the original transpose routines from \texttt{2DECOMP\&FFT}.
\par The discrete transforms are computed with the \texttt{FFTW3} library \cite{FFTW3}. Apart from its inherent efficiency, the \emph{Guru} interface of the library computes all the discrete transforms presented in table~\ref{tbl:transfs} with the same syntax, just by modifying a parameter corresponding to the transform type\footnote{Note that  the result of a \emph{real-to-complex} DFT of $n$ real numbers only requires storage of $n$ real numbers, and therefore can be handled in the same way as the other DT.}. This allowed for a single and efficient parallel implementation of the Poisson solver, as only one data layout for the transposing routines is required. The resulting tridiagonal system of equations is solved with the \texttt{LAPACK} library (\texttt{DGTSV}) \cite{Lapack}. In case of periodicity in $z$, the resulting cyclic tridiagonal system is reduced into two tridiagonal problems; see e.g.\ \cite{BookNA}.\par
The algorithm for solving the Poisson equation is summarized below:
\begin{itemize}
    \item[1:] compute the RHS of the Poisson equation in the $z$-aligned pencil decomposition, and transpose result to $x$-aligned pencil decomposition;
    \item[2:] compute $N_y N_z$ forward 1D {DT} in $x$, and transpose result to $y$-aligned pencil decomposition;
    \item[3:] compute $N_x N_z$ forward 1D {DT} in $y$, and transpose result to $z$-aligned pencil decomposition;
    \item[4:] solve $N_x N_y$ linear tridiagonal systems with Gauss elimination, and transpose result to $y$-aligned pencil decomposition;
    \item[5:] compute $N_x N_z$ backward 1D {DT} in $y$, and transpose result to $x$-aligned pencil decomposition;
    \item[6:] compute $N_y N_z$ backward 1D {DT} in $x$, and transpose result to $z$-aligned pencil decomposition.
\end{itemize}
Finally, parallel I/O is also handled by the \texttt{2DECOMP\&FFT}, which is based on \texttt{MPI-I/O}.

\section{Validation and Computational performance}
\label{sec:results}
\subsection{Validation}
We validated our implementation against three wall-bounded flows, with different combinations of boundary conditions, as shown in table~\ref{tbl:simus}. Hereafter, $u$, $v$ and $w$ denote the $x$-, $y$- and $z$-components of the mean velocity, respectively.
\begin{table}
\centering
\caption{Physical and computational parameters for the validation cases. $L_{x_i}$ and $N_{x_i}$ denote the domain size and number of points in direction $x_{i}$, respectively. See the text for the scaling parameters used for the Reynolds number, $\mathrm{Re}$.}\label{tbl:simus}
\begin{tabular}{ l c c c c }
 \hline\hline
 Case                & $ L_x \times L_y  \times L_z  $ & $ N_x \times N_y \times N_z$  & Pressure BC in x,y,z   & $\mathrm{Re}$ \\
 \hline\hline                                                                                                       
 lid-driven cavity   & $ ~~1 \times 1~   \times 1    $ & $ 128\times 128 \times 128$ & \texttt{N-N, N-N, N-N} & $1000$        \\
 \hline                                                                                                             
 square duct         & $ 10 \times  1~   \times 1    $ & $ 512\times 128 \times 128$ & \texttt{P-P, N-N, N-N} & $4410$        \\
 \hline                                                                                                          
 plane channel       & $ ~~6 \times 3~   \times 1    $ & $ 512\times 256 \times 144$ & \texttt{P-P, P-P, N-N} & $5640$        \\
 \hline
    Taylor-Green vortex & $ 2\pi\times 2\pi \times 2\pi $ & $ 512\times 512 \times 512$ & \texttt{P-P, P-P, P-P} & $1600$       \\
 \hline                                                                                                    
 \hline
\end{tabular}
\end{table}
\subsubsection{Lid-driven cavity flow}
We simulated a lid-driven cavity flow in a cubic domain with dimensions $[-h/2,h/2]^3$. Zero-velocity no-slip and no-penetration boundary conditions are prescribed at all the boundaries, except for the top wall, which moves with a velocity $\mathbf{u}(x,y,h/2)=(U_L,0,0)$. Other physical and computational parameters are shown in table~\ref{tbl:simus}, where the Reynolds number is defined as $\mathrm{Re}= U_Lh/\nu$.\par
\begin{figure}[!htbp]
  \centering
  \includegraphics[width=0.49\textwidth]{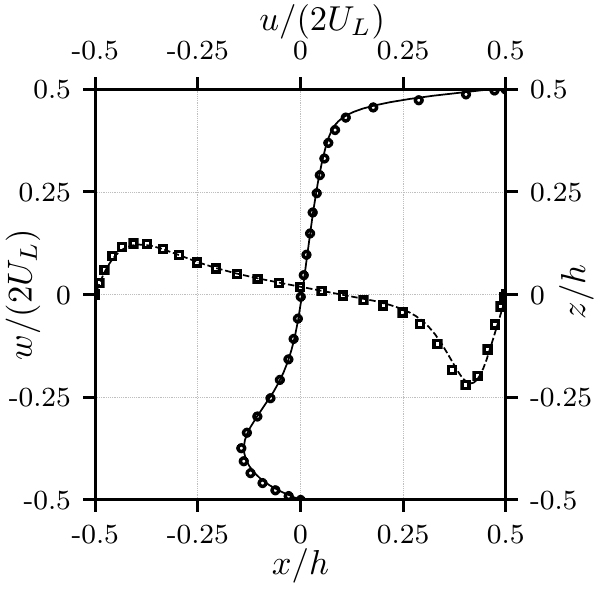}
  \caption{Normal velocity profiles along the centrelines $u(0,0,z)$ and $w(x,0,0)$ for a lid-driven cubic cavity at $\mathrm{Re}=1000$. The symbols correspond to DNS data extracted from \cite{Ku-et-al-JCP-1987}.}\label{fig:val_ldc}
\end{figure}

Figure~\ref{fig:val_ldc} shows the velocity profiles of the steady state solution at the centrelines $u(0,0,z)$ and $w(x,0,0)$, compared to the data extracted from \citet{Ku-et-al-JCP-1987}. The results show excellent agreement with the data.

\subsubsection{Pressure-driven turbulent channel and square duct flow}
We now consider two turbulent wall-bounded flows: a plane channel and a square duct. Both flows are periodic in the streamwise ($x$) direction, with no-slip/no-penetration boundary conditions at the wall-normal directions ($y=\pm h$ and $z=\pm h$ in the case of the duct, and $z=\pm h$, with periodicity in the spanwise direction $y$, in the channel case). A volume force is added to the discretized momentum equation, to maintain a bulk streamwise velocity $U_b=1$. The physical and computational parameters are shown in table~\ref{tbl:simus}, where $\mathrm{Re}= U_b(2h)/\nu$, and $h$ is the channel/duct half-height.
\par
\begin{figure}[!htbp]
  \includegraphics[width=0.99\textwidth]{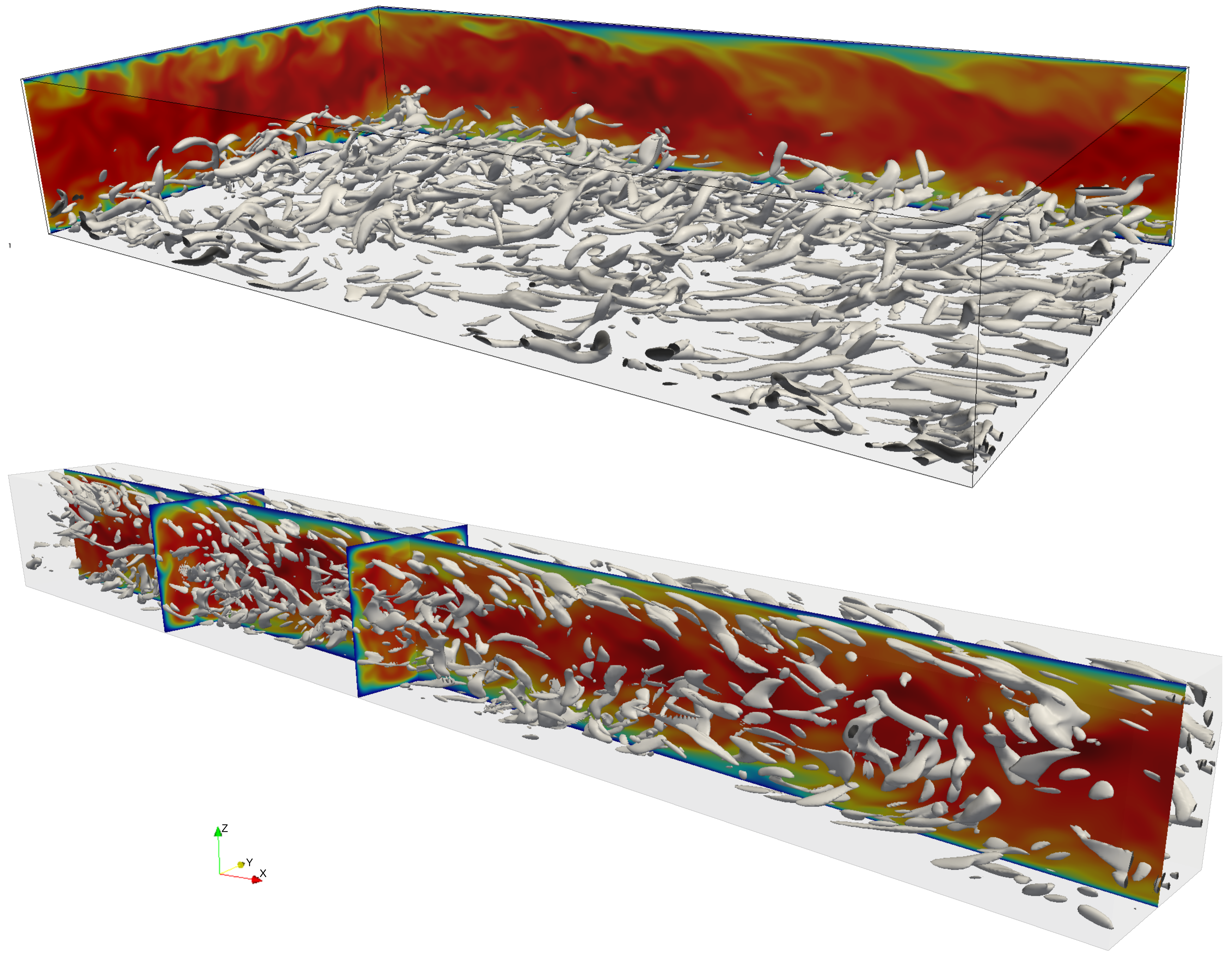}
	\caption{Visualizations of the simulations of the turbulent channel (top) and duct (bottom). Q-criterion shown with iso-contours of the second invariant of the velocity-gradient tensor, $Q/(U_b/(2h))^2=5$ (for the channel only the lower half is shown). The contours pertain to streamwise velocity (red-- high, blue-- low).}\label{fig:visuchan}
\end{figure}

 Figure~\ref{fig:visuchan} shows a 3D visualization for the two flow cases, illustrating what is typically seen for a turbulent flow at low Reynolds number: three-dimensional coherent structures with a relatively small scale-separation with respect to the scales of confinement.\par 
The channel flow simulation was initialized with a streamwise vortex pair \cite{Henningson-and-Kim-JFM-1991} which effectively triggered transition. Statistics were collected once the mean pressure gradient required to sustain the constant flow rate reached a statistically steady state. The data were ensemble-averaged from $1000$ samples, over a period of $1400 h/U_b$.\par
\begin{figure}[!htbp]
  \includegraphics[width=0.49\textwidth]{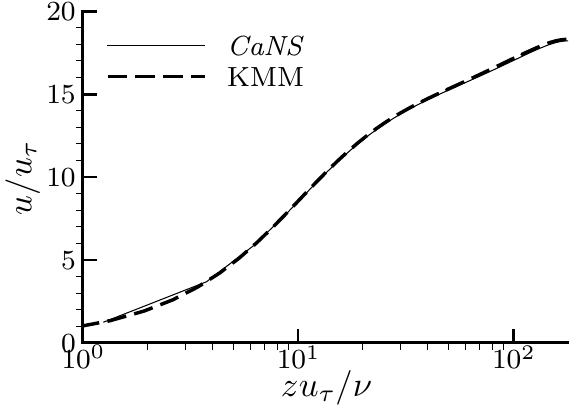} \hfill
  \includegraphics[width=0.49\textwidth]{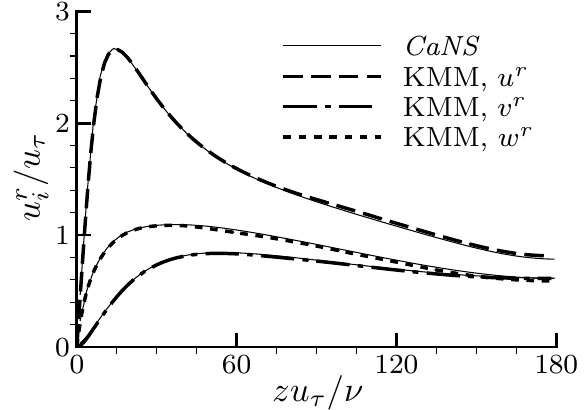}
  \caption{Left: mean streamwise velocity profile for turbulent channel flow at friction Reynolds number $\mathrm{Re}_\tau=180$. Right: profiles of root-mean-square velocity $u_i^r$. Both figures use inner-scaling, i.e.\ velocity scaled with the wall friction velocity $u_\tau$, and distance with the viscous wall-unit $\nu/u_\tau$. The profiles are compared to DNS data from \cite{Kim-et-al-JFM-1987} (KMM).}\label{fig:valchan}
\end{figure}

We computed the friction Reynolds number from the time-averaged pressure gradient, yielding $\mathrm{Re}_\tau=u_\tau h/\nu=180.2$, with $u_\tau\equiv\sqrt{-(\mathrm{d}p_m/\mathrm{dx})h}$ the wall-friction velocity. This agrees with the correlation given by \citet{Pope}: $\mathrm{Re}_\tau=0.09\mathrm{Re}^{0.88}\approx 180$. Figure~\ref{fig:valchan} compares first and second-order mean flow statistics against data from the seminal paper of \citet{Kim-et-al-JFM-1987} at the same friction Reynolds number. Again, the results closely match the reference DNS data.
\par
Now for the turbulent square duct, we used an initial condition from a DNS performed in our group at similar Reynolds number, which allows for a fast transient towards a fully-developed turbulent state. Since the mean flow is two-dimensional, statistical convergence of the results requires many samples. The data were averaged from $1000$ samples, over a period of $15000 h/U_b$.\par
\begin{figure}[!htbp]
  \includegraphics[width=0.49\textwidth]{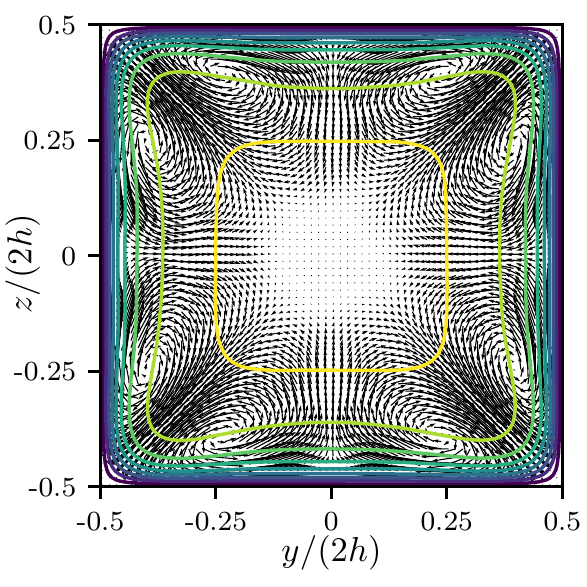}\hfill
  \includegraphics[width=0.49\textwidth]{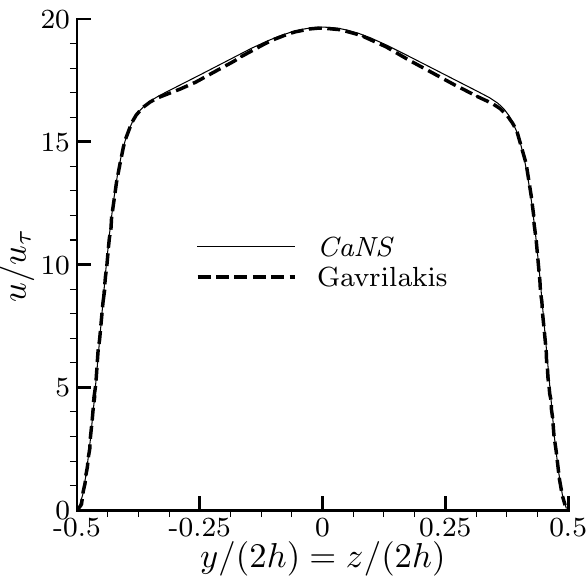}
  \put(-180,0){$(a)$}
  \put( -15,0){$(b)$}
  \caption{(a): mean flow for a square duct. The isolines pertain to streamwise velocity, starting with $2u_\tau$, and evenly-spaced by the same amount. The vectors illustrate in-plane velocity, with a maximum magnitude of $0.02 U_b$. (b): mean streamwise velocity along the duct diagonal, compared to the data extracted from \citet{Gavrilakis-JFM-1992}.}\label{fig:valduct}
\end{figure}

Like for the turbulent channel, we compute the friction Reynolds number based on the wall-friction velocity, obtaining a value of $\mathrm{Re}_\tau=149.1$, consistent with the value of $\approx 150$ reported by \citet{Gavrilakis-JFM-1992} for a DNS with the same parameters. Figure~\ref{fig:valduct} (a) quantifies the mean flow. It is well-known that the presence of corners induces a non-zero (secondary) mean flow in the wall-normal directions, in this case with a maximum magnitude of about $2\%$ of the bulk velocity. This value is also consistent with the results reported by \citet{Gavrilakis-JFM-1992}. Panel (b) of figure~\ref{fig:valduct} compares the streamwise velocity profile along the diagonal of the duct cross-section, compared to the data extracted from \cite{Gavrilakis-JFM-1992}. The results show good agreement.

\subsubsection{Taylor-Green vortex}
The last validation considers the temporal evolution of a Taylor-Green vortex. The flow is solved in a tri-periodic domain with dimensions $[0,2\pi]^3$, with the following initial condition for the velocity field $\mathbf{u}(\mathbf{x},t=0)=(u_0,v_0,w_0)$:
\begin{align}
u_0 &= ~~U\sin(x/L)\cos(y/L)\cos(z/L)  \mathrm{,} \label{eqn:tgu} \\
v_0 &= - U\cos(x/L)\sin(y/L)\cos(z/L)  \mathrm{,} \label{eqn:tgv} \\
w_0 &= 0                      \,\mathrm{.} \label{eqn:tgw}
\end{align}
with $U=1$, $L=1$, and a Reynolds number $\mathrm{Re}_{TG} \equiv UL/\nu = 1600$. Other computational parameters are shown in table~\ref{tbl:simus}.\par
\begin{figure}
	\centering
  \includegraphics[width=0.99\textwidth]{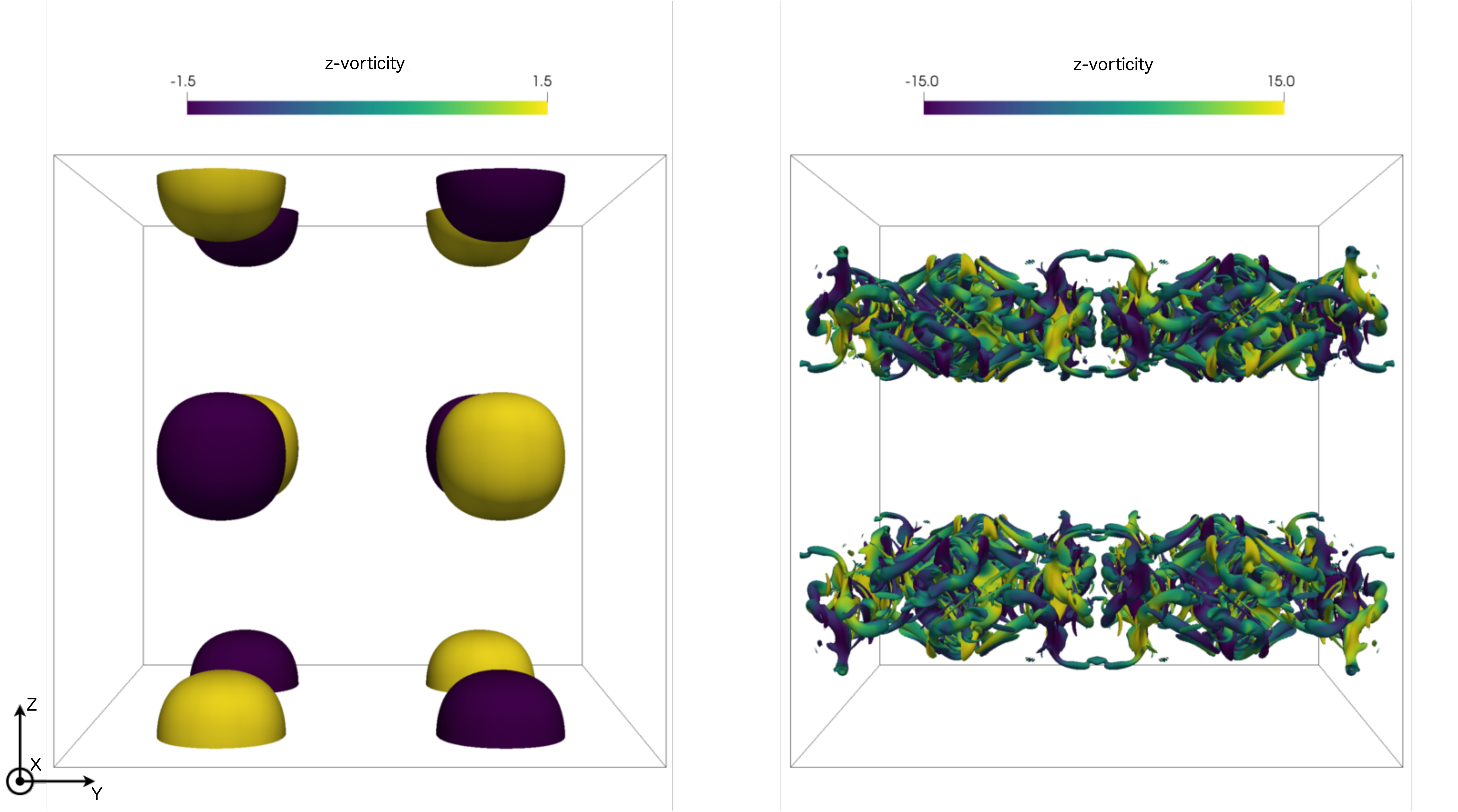} \\
  \includegraphics[width=0.49\textwidth]{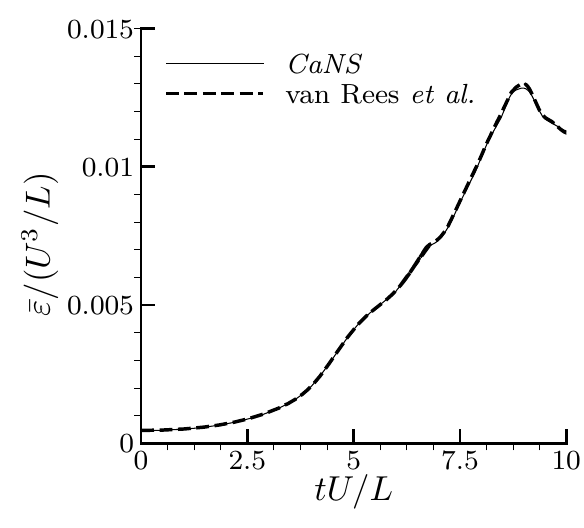}
  \put(  80,150){$(a)$}
  \put(  10,10){$(b)$}
    \caption{(a): visualization of iso-surfaces of vorticity magnitude $|\boldsymbol{\Omega}|$ for the Taylor-Green vortex benchmark. The left panel shows the initial condition with $|\boldsymbol{\Omega}|=1.5U/L$ and the right panel with $|\boldsymbol{\Omega}|=15 U/L$ at instant $t=9L/U$. The colors correspond to the $z$-component of vorticity (see the legend). (b): temporal evolution of the mean viscous dissipation $\bar{\varepsilon}$ of a three-dimensional Taylor-Green vortex, compared to the
    reference data from a pseudo-spectral code in \cite{vanRees-et-al-JCP-2011}.}\label{fig:tgv}
\end{figure}

In this case, a smooth initial velocity field will produce vorticity due to vortex-stretching, generating small-scale vortical structures \cite{Brachet-et-al-JFM-1983}. This mechanism can be visualized in Figure~\ref{fig:tgv}(a), which shows iso-contours of vorticity magnitude for the initial condition and the instant corresponding to the maximum value of energy dissipation.
\par
The generation of small-scale vortical structures leads to a net increase in viscous dissipation of kinetic energy, $\varepsilon = 2\nu \mathbf{S}\cddot\mathbf{S}$; with $\mathbf{S}$ being the strain-rate tensor $\mathbf{S}\equiv (\boldsymbol{\nabla}\mathbf{u}+\boldsymbol{\nabla}\mathbf{u}^T)/2$. For larger times, this quantity will show a net decrease and eventually vanish, since there is no external power input. Figure~\ref{fig:tgv}(b) shows the expected trend, with the mean (i.e.\ space-averaged)
viscous dissipation reaching a maximum at $t=9L/U$. The results are compared to the reference data in \cite{vanRees-et-al-JCP-2011}, showing good agreement.
\subsection{Computational performance}
Finally, we test the scaling performance of our implementation for a tri-periodic domain. Since the DFT is cheaper than the other DT, this case corresponds to the highest memory bandwidth per FLOP. It should be, therefore, the worst case scenario for a scaling test.\par
The simulations were performed in partition A2 (Knights Landing) of the supercomputer MARCONI from Cineca, Italy. Only distributed-memory parallelization was tested in the present work, and sufficed for achieving good scaling. The shared-memory implementation may be useful for future extensions, but as implemented now does not seem to improve the performance.\par
\begin{figure}[!htbp]
  \includegraphics[width=0.49\textwidth]{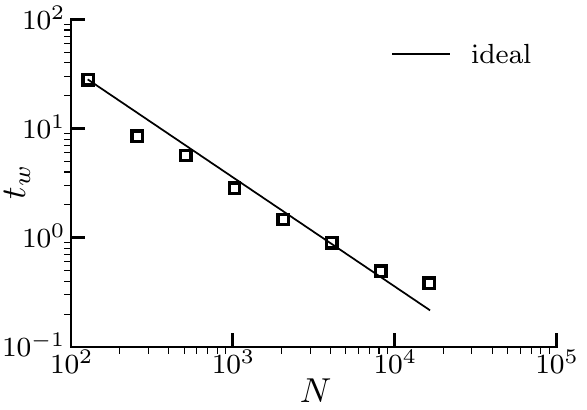}\hfill
  \includegraphics[width=0.49\textwidth]{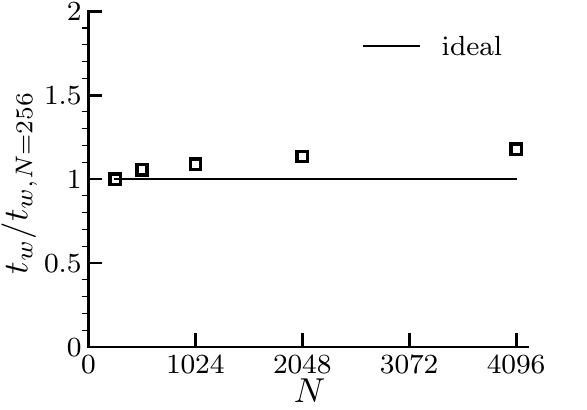}
  \put(-180,0){$(a)$}
  \put( -15,0){$(b)$}
  \caption{(a): strong scaling of the numerical method up to $16384$ cores in a domain with $1024^3$ grid cells. $t_{w}$ denotes wall-clock time in seconds, and $N$ the number of cores. (b): weak scaling performance for a domain with $1024^3/512\approx 2\cdot 10^6$ grid cells per core. $t_{w}$ denotes mean wall-clock time in seconds/time step/task (i.e.\ three Runge-Kutta substeps), and $N$ the number of cores. $t_{w,256}$ corresponds to the wall-clock time for $N=256$.}\label{fig:scaling}
\end{figure}

 
The strong scaling tests were performed in a domain with $1024^3$ grid cells. Figure~\ref{fig:scaling}~(a) shows the wall-clock time versus the number of cores. Superlinear speedup can be noticed for a smaller number of cores, likely due to cache effects. Overall, the code shows very good scaling performance up to about $O(10^4)$ cores, reaching a small wall-clock time $t_{w}\approx 0.5\mathrm{s}/\mathrm{core}/\mathrm{time step}$. These numbers are consistent with the good performance of the \texttt{2DECOMP\&FFT} library \cite{Li-and-Laizet-CRAY-2010}, which handles the most demanding parallelization steps.\par
Weak scaling tests are illustrated in figure~\ref{fig:scaling}~(b), with the number of grid cells per task fixed to $2\cdot 10^6$. The results show a slight monotonic deterioration of up to $17\%$ from $N=256$ to $4096$ cores. Overall, the results indicate that strong scaling timings are likely to scale up to much larger problem sizes.
\section{Conclusions and outlook}\label{sec:conclusions}
We presented an efficient numerical algorithm for massively-parallel DNS of canonical turbulent flows. The method uses a direct, FFT-based solver for the pressure Poisson equation discretized with second-order central differences, parallelized with a 2D \emph{pencil}-like domain decomposition. This approach has been applied recently to massively-parallel numerical simulations of complex turbulent flows with $O(10^9)$ spatial degrees of freedom, but restricted to at least two periodic directions; see e.g.\ \cite{Dodd-and-Ferrante-JFM-2016,Costa-et-al-PRL-2016,Ostilla-et-al-JFM-2016}. To the best of our knowledge, this is the first general implementation of such parallel algorithms allowing for the different combinations of homogeneous pressure boundary conditions, that can benefit from the method of eigenfunction expansions. Our approach was shown to scale up to about $10^4$ cores for a problem with $10^9$ spatial degrees of freedom, reaching a very small wall-clock time. These figures will probably scale for larger problem sizes.\par
The method was validated against distinct benchmark cases of canonical laminar and turbulent flows. It should be noted that several other configurations could have been considered. Obvious examples are wall-bounded flows with inflow/outflow boundary conditions.
\par
For low Reynolds number flows (or extremely high resolution), implicit temporal integration of the diffusion term can be advantageous. In that case, non-staggered discrete transform operators can be considered for non-periodic cases, but the velocity boundary conditions in $x$ and $y$ must be homogeneous. This has been implemented in our numerical tool.\par
This type of Navier-Stokes solvers, combined with other methodologies to handle, e.g., complex geometries or multiphase flows, have been unveiling important physical insights into flows that require massively-parallel DNS. In the same spirit, the resulting open-source code, more than a tool for simulations of canonical flows, can be seen as an efficient base solver on top of which numerical methods for more complex flows can be implemented.
\section*{Acknowledgments}
Wim-Paul Breugem and Bendiks Jan Boersma from the TU Delft are acknowledged
for interesting discussions. Also, the structure of the open-source DNS code has been
inspired by the in-house DNS codes used in their groups. Vilborg Gudj\'onsd\'ottir from TU Delft is thanked for proof-reading this manuscript.\par
Marco Rosti and Mehdi Niazi from KTH Mechanics are thanked for providing the initial conditions for the turbulent duct and channel flow simulations, respectively.\par
The implementation ran on three different systems. We acknowledge the computing time in the Dutch National Supercomputer Cartesius granted by NWO (Netherlands Organisation for Scientific Research), for first development tests. The validation simulations were performed on the supercomputer Beskow at the PDC center, KTH, with computing time provided by SNIC (Swedish National Infrastructure for Computing). Finally, the supercomptuer MARCONI from Cineca in Italy was used under a PRACE grant for the scaling simulations.


\bibliography{bibfile}

\end{document}